\documentstyle{article}
\begin{document}

\begin{titlepage}

\begin{flushright}
\large
INP MSU Preprint-95-2/366\\
Journal-ref: Int. J. of Mod. Phys. C7, No.6 (1996) pp.761-774 \\
18 December 1996
\end{flushright}

\vskip 2cm
\begin{center}
{\large\bf\boldmath{
{RECURSIVE ALGORITHM FOR THE GENERATION OF 
RELATIVISTIC KINEMATICS FOR COLLISIONS AND DECAYS
WITH REGULARIZATIONS OF SHARP PEAKS}
}}
\end{center}
\bigskip
\begin{center}
\large
{V.A.~Ilyin, D.N.~Kovalenko, A.E.~Pukhov \\
{\small\it Institute of Nuclear Physics of Moscow State University,
	119899 Moscow, Russia}
 }
\end{center}\bigskip

\vskip 3cm
\begin{abstract}
\noindent
\large
We describe an algorithm for the generation of
relativistic kinematics for collision and decay processes with
multiparticle final states. In the framework of this algorithm
it is possible to generate different kinematics covering most of
practically interesting cases. One gets a possibility to
introduce different sets of integration variables. As a result
different sets of kinematical singularities can be regularized.
To smooth sharp peaks some regularization formulas and
procedures are used covering most typical cases. The algorithm
is realized in the package CompHEP created for automatic
calculations of collision and decay processes.
\end{abstract}

\end{titlepage}

\section{Introduction}        
\noindent
To evaluate numerically cross sections and other characteristics
of elementary particle collisions and to calculate decays of
unstable particles it is necessary to integrate squared matrix
element over the phase space.

The first step is to express the integrand in terms of
independent kinematical variables --- integration variables.
Note that after evaluation of Feynman diagrams the integrand is
expressed in terms of scalar products of Lorentz momenta of {\it
in-\/} and {\it out-\/} particles, and polarization vectors.

The next step is connected with a singular behavior of the
integrand that is typical case in modern collider physics
calculations (see Ref.~\cite{1}, for example).  In many cases
straightforward methods of the numerical integration fail
because of sharp peaks.  Standard procedure to treat carefully
such peaks is to transform integration variables in order to get
nonsingular function in the new variables. We will call such
procedure the {\it regularization.}

These two steps are closely connected. To make it clear let us
note that  singularities appear as zeros (more
exactly --- almost zeros) of propagators denominators. So one
can say that each singularity is connected directly with a
squared Lorentz 4-vector expressed in its order in terms of
linear combination of particle momenta.  However it is clear
that singularity can be smoothed by integration
variable transformation only if the corresponding invariant is
one of these variables or related directly to such variable.
Thus one can integrate carefully if both steps will be performed
in proper relation to the set of singularities. 
This problem becomes nontrivial especially in so-called automatic
calculation of collision and decay processes. This new approach,
based on high perfomance and user-friendly programs, is
developed now intensively Refs.~\cite{2}-\cite{4}. With the help
of these programs one can calculate different processes in the
framework of the various physical models. It is clear that the
generation of kinematics is an important part of this approach.

There are many known kinematical schemes implemented in existing
approaches (see Ref.~\cite{5}, for example).  In
GRACE \cite{3} special library of kinematics, corresponded to
different sets of kinematical singularities, was created.  In
EXCALIBUR \cite{6} a multichannel appproach to compute four fermion
processes in $e^+e^-$ annihilation is presented.  In this paper
we demonstrate that the problem discussed can be treated in the
framework of some general algorithm based on the recursive
representation of kinematics as a tree of elementary decays
$1\rightarrow 2$.  It opens the possibility to generate
kinematics (choose integration variables, express squared matrix
elements in terms of these variables and introduce proper
regularizations) for almost all kinds of processes with the help
of one program with a high level of user's interface. This
algorithm was realized in the package CompHEP \cite{2} and
representative computational examples are given here.

\section{Kinematics}
\noindent
By definition, the kinematical scheme is a way
of phase space parameterization.

We define the kinematical scheme as a tree of {\it elementary
decays.} Under the {\it elementary decay\/} we understand the
parameterization in 2-particle phase space.  Note that the words
{\it elementary decay\/} do not correspond to real $1\rightarrow
2$ decay in most cases.  We use this terminology for
convenience.

Below we consider multiparticle phase space for collisions. For
decays the corresponding formulas can be written in a similar
way.

Let us take any two out-particles with the numbers $i$ and $j$.
We consider their momenta as momenta produced in
the following {\it elementary decay\/}:
\begin{equation}	
q_{ij}\rightarrow q_i\,, q_j\,,\label{eq:decay}
\end{equation}
where $q_{ij}=q_i+q_j=(E_{ij},\, \vec{q}_{ij})$.  Let us
introduce a parameterization for this {\it elementary
decay\/} in the rest frame of decayed 4-vector $q_{ij}$ as
\begin{equation}	
d\Gamma_2(i,\, j)\equiv
\frac{R_i}{(2\pi)^3\cdot 4\sqrt{s_{ij}}}\cdot d s_{ij}\cdot
d\Omega_i\,.\label{eq:gamm2}
\end{equation}
Here the measure $d\Omega_i$ corresponds to the angular part of
$d\vec{q}_i$.  Then $R_i=|\vec{q}_i|$ in the rest frame of
$q_{ij}$, and $s_{ij}=q_{ij}^2.$

We repeat now this construction recursively: on each step two
cluster 4-vectors $q_I$,$q_J$ are taken from the rest part of
out-momenta or from the decayed 4-vectors $q_{I\!J}$ appeared on
previous steps. After $(N-1)$ steps we get the following formula
for differential cross section ($|M|^2$ is a squared matrix
element --- the normalization corresponds to the
conventions \cite{7}):
\begin{equation}	
d\sigma=\frac{\pi}{2 R_1\sqrt s} |M|^2\cdot
d\Gamma^*_2(I_1,\, J_1)\cdot
\prod_{i=2}^{N-1}d\Gamma_2(I_i,\, J_i)\,,
\end{equation}
where the 2-particle phase space measure for the 1st {\it
elementary decay\/} $p_1+p_2\rightarrow q_{I_1}$, $q_{J_1}$ does
not include integration over $s\equiv (p_1+p_2)^2$ (here $p_1$
and $p_2$ are momenta of colliding particles):
\begin{equation}	
d\Gamma^*_2(I_1,\, J_1)\equiv
\frac{R_{I_1}}{(2\pi)^3\cdot 4\sqrt s}\cdot d\Omega_{I_1}\,.
\label{eq:gamm3}
\end{equation}
Here $R_{I_1}\equiv\{|\vec{q}_{I_1}|,\, \mbox{\it in CMS}\}$, by
CMS we denote the kinematical frame of center mass system.

The angular measure $d\Omega_I$ can be written in terms of two
angular coordinates of the 3-vector $\vec{q}_I$. Let us
introduce two reference 4-vectors 
${\cal P}_I=({\cal P}^0_I,{\bar {\cal P}}_I)$ 
and ${\cal A}_I=({\cal A}^0_I,{\bar {\cal A}}_I)$
for some {\it elementary decay\/} in the rest frame of the
decayed vector $q_{I\!J}$.  So, let the polar coordinate of
$\vec{q}_I$, the angle $\Theta_I$, be counted from the 3-vector
${\bar {\cal P}}_I$.  Then let us take the azimuth coordinate
of $\vec{q}_I$ as the angle $\Phi_I$ between two planes
(${\bar {\cal P}}_I,\, {\bar {\cal A}}_I$) and
(${\bar {\cal P}}_I,\, \vec{q}_I$). As a result the measure
$d\Omega_I$ takes the form
\begin{equation}	
d\Omega_I=d\cos\Theta_I\cdot d\Phi_I\,.
\label{eq:dtheta}
\end{equation}

Taking the reference vectors as different combination of the
momenta of in- and out-particles we can connect different
Lorentz invariants with introduced integration variables. Let us
consider, for example, {\it elementary decay\/}
$q_{I\!J}\rightarrow q_I$, $q_J$ and ${\cal P}_I=\sum_n p_n$,
where $p_n$ are momenta of some in- and out-particles.  The relation 
between the angle $\Theta_I$ and the invariant $(\sum p_n+q_I)^2$ can 
be written in the following form:
\begin{equation}	
\left(\sum p_n+q_I\right)^2=\left(\sum p_n\right)^2+(q_I)^2
+2\left(\sum p_n^0\right)\cdot q_I^0
-2\left|\sum\vec{p}_n\right|\cdot|\vec{q}_I|\cos\Theta_I\,.
\label{eq:varrel}
\end{equation}
In the following we use this relation to introduce
regularizations of sharp peaks\footnote{The relation between 
angle $\Phi_I$ and some Lorentz invariant
is more complicated because it includes also $\cos\Theta$. Of
course, it is possible to use the angle $\Phi_I$ for the
regularization purposes. However, we tested such regularizations
and found them to be not effective and we do not consider this
possibility here. For integration over
$\Phi_I$ we take some
fixed reference vector ${\bar {\cal A}}_I$.  It could be unit vector
orthogonal to the collision axes, for example.}.

Let us introduce numbering of the {\it elementary decays} by the following
rules:
\begin{enumerate}
\item initial {\it elementary decay} $(p_1+p_2\to q_{I_1},q_{J_1})$ 
has a number $i=1$;
\item all {\it elementary decays} on branches, growing from the clusters
$I_i$ and $J_i$ of the $i$-th {\it elementary decay},
have numbers larger than $i$;
\item any {\it elementary decay} on the branch growing from the cluster
$I_i$ has a number smaller than any {\it elementary decay} on the branch
growing from the second cluster $J_i$ of the $i$-th {\it elementary decay}.
\end{enumerate}
We will use a terminology `previous {\it elementary decay}' if it has 
smaller number than the current one.
In CompHEP realization the integrations in (3) are ordered in the
correspondence with this numbering of {\it elementary decays}. 

\subsection{Kinematics summary}
\vspace*{-3pt}
\begin{itemize}
\item The kinematical scheme is a {\it tree of ``elementary decays".}
\item The number of {\it elementary decays\/} equals $(N_{\rm
      out}-1)$, where $N_{\rm out}$ is a number of out-particles.
\item Each {\it elementary decay\/} is defined by
  \begin{itemize}
  \item three 4-vectors: decayed $q_{I\!J}$ and two cluster vectors,
    $q_I$ and $q_J$, with $q_{I\!J}=q_I+q_J$;
  \item reference vector ${\cal P}_I$ 
    which can be chosen from some combinations of particle momenta.
   These combinations have to be already defined as cluster momenta
   or decayed vectors in previous {\it elementary decays.}
  \end{itemize}
\item The vectors $q_{I\!J}$ and
   ${\cal P}_I$ has to be linearly independent.
\item Each {\it elementary decay\/} $q_{I\!J}\to q_I$, $q_J$ adds
   three new integration variables:
  \begin{itemize}
  \item $s_{I\!J}$ --- squared decayed vector $q_{I\!J}$;
  \item $\Theta_I$ and $\Phi_I$ --- polar and azimuth angles of
   $\vec{q}_I$ counted from the reference 3-vectors ${\bar {\cal P}}_I$
   and ${\bar {\cal A}}_I$ in the rest frame of $q_{I\!J}$.
   For decays or collisions with fixed energies of in-particles the
   1st {\it elementary decay\/} gives only two independent
   variables, $\Theta_1$ and $\Phi_1$.  In that case the number of
   integration variables equals $3\cdot(N_{\rm out}-1)-1=3\cdot
   N_{\rm out}-4$.
  \end{itemize}
\end{itemize}

\vspace*{-5pt}
\section{Invariant Cuts}
\noindent
In the framework of the algorithm described above it is possible
to introduce effectively kinematical cuts over some combinations
of Lorentz momenta of in- and out-particles.

In general we can introduce an invariant cut in two ways.  The
first is more effective and it can be realized if the
corresponding invariant is connected directly with one of the
integration variables of kinematical scheme (e.g., as
Eq.~(\ref{eq:varrel})).  In this case the corresponding cut can
be implemented by the limitation of the integration volume.

In other cases entered cuts can be applied by multiplication of
squared matrix element on the corresponding step function.  This
way can be not so effective, especially in the case of strong
cut when the integrand equals zero in almost all phase space
volume.

\section{Regularization}
\noindent
In some cases the squared matrix element has singularities when
the integrand has sharp peak(s) in phase space and the standard
Monte Carlo integrator (even adaptive one) could fail.  The
typical examples are given below.

\subsection{General regularization formula}
\noindent
Let the integrand $F(x)$ to be represented as
$F(x)=f(x)\cdot(g_1(x)+\cdots+g_n(x))$, where $f(x)$ is a
smooth function and functions $g_i(x)$ have sharp peaks.  One
can transform the integration variable in the following way:
\begin{equation}	
dx=\frac1{\displaystyle\sum_{i=1}^ng_i(x(\tilde y))}d\tilde y\,,
\qquad
\tilde y(x)=\sum_{i=1}^nG_i(x)\equiv G(x)\,,
\end{equation}
where integral functions are $G_i(x)=\int^x_a g_i(x')dx'$.
The function
$x(\tilde y)\equiv G^{-1}(\tilde y)$ we will find from the relation
$\sum_{i=1}^nG_i(x)-\tilde y=0$ numerically.
Also let us transform the integration volume to the interval
$[0,\, 1]$: $d\tilde y=(G(b)-G(a))dy$, $0\leq y\leq 1\,.$

As a result the integral can be rewritten as
\begin{equation}	
\int_a^b F(x)dx=\int_0^1f(x(y))\cdot\left(\sum_{i=1}^ng_i(x(y))\right)
J_{\{g\}}(y)dy\,,
\end{equation}
where Jacobian equals $J_{\{g\}}(y)=\frac1{\sum_{i=1}^n
g_i(x(y))} (G(b)-G(a))$.

We see that the singular integral is transformed now to smooth
integration of the function $f(x)$. Of course integrand can have
more complicated functional structure then we have considered.
However, if we know the structure of singularities it is not so
difficult to write some model function of the form
$(g_1(x)+\cdots+g_n(x))$ which will match each singularity. And
as a result the transformation of the integration variable
described above will regularize the integrand.  Here we would
like to stress again that the structure of singularities can be
derived easily from the analysis of denominators of calculated
Feynman diagrams.

Note that it is possible to smooth several peaks over one
variable appearing from the propagators of different Feynman
diagrams. In this case the model singular function
$\sum_{i=1}^ng_i(x)$ is constructed with $g_i(x)$ corresponding
to separate peaks.

Then we consider the constant $g_n(x)=1$ as one of
model function. It is necessary to avoid neglect of contributions
of nonsingular areas after regularization.

\subsection{Types of regularizations}

\subsubsection{$t$-channel pole}
\begin{figure}
\linethickness{0.5pt}
\begin{center}
\begin{picture}(62,77)(0,0)
\put(17.5,57.1){\vector(1,0){2}}
\put(9.9,57.1){\makebox(0,0)[r]{$e^-$}}
\put(10.9,57.1){\line(1,0){13.0}}
\put(30.2,60.2){\makebox(0,0){$Z$}}
\multiput(23.7,57.1)(0.3,0.0){12}{\rule[-0.25pt]{0.5pt}{0.5pt}}
\multiput(28.6,57.1)(0.3,0.0){12}{\rule[-0.25pt]{0.5pt}{0.5pt}}
\multiput(33.5,57.1)(0.3,0.0){12}{\rule[-0.25pt]{0.5pt}{0.5pt}}
\put(51.1,70.2){\makebox(0,0)[l]{$Z$}}
\multiput(36.7,57.1)(0.3,0.3){12}{\rule[-0.25pt]{0.5pt}{0.5pt}}
\multiput(41.6,62.0)(0.3,0.3){12}{\rule[-0.25pt]{0.5pt}{0.5pt}}
\multiput(46.5,66.9)(0.3,0.3){12}{\rule[-0.25pt]{0.5pt}{0.5pt}}
\put(51.1,44.0){\makebox(0,0)[l]{$H$}}
\multiput(36.5,57.1)(3.3,-3.3){5}{\rule[-0.5pt]{1.0pt}{1.0pt}}
\put(24.0,44.0){\vector(0,-1){2}}
\put(21.9,44.0){\makebox(0,0)[r]{$e^-$}}
\put(24.0,57.1){\line(0,-1){26.2}}
\put(9.9,30.9){\makebox(0,0)[r]{$\gamma$}}
\multiput(10.7,30.9)(0.3,0.0){12}{\rule[-0.25pt]{0.5pt}{0.5pt}}
\multiput(15.6,30.9)(0.3,0.0){12}{\rule[-0.25pt]{0.5pt}{0.5pt}}
\multiput(20.5,30.9)(0.3,0.0){12}{\rule[-0.25pt]{0.5pt}{0.5pt}}
\put(24.0,30.9){\line(1,0){13.0}}
\put(43.5,24.4){\vector(1,-1){2}}
\put(51.1,17.8){\makebox(0,0)[l]{$e^-$}}
\put(37.0,30.9){\line(1,-1){13.0}}
\end{picture}
\end{center}
\caption{Feynman diagram of $\gamma+e^-\to e^-+Z+H$ with
$t$-channel type of singularity, produced by the exchange of
virtual electron.}\label{fig:diag1}
\end{figure}
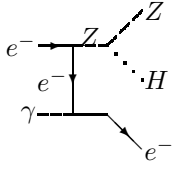

\noindent
Let us consider a typical example when in some Feynman diagram
the in-particle with momentum $p_1$ radiates the virtual
particle with mass $m$ and after that transforms to the
out-particle with momentum $p_3$.  The corresponding propagator
of the virtual particle is
\begin{equation}	
\frac1{t-m^2}\,,\quad t\equiv (p_3-p_1)^2\,.\label{eq:tprop}
\end{equation}
The integrand will have sharp peak in $t$ near its upper bound 
$t_{\max}$ if $m^2$ and
$t_{\max}$ are much less than the energy of in-particles:
\begin{equation}	
m^2\ll s\,,\quad t_{\max}\ll s\,.
\end{equation}
Typical example is given in Fig.~\ref{fig:diag1} by the Feynman
diagram of the process
$$\gamma(p_1)+e^-(p_2)\rightarrow e^-(p_3)+Z(p_4)+H(p_5)\,,$$
where $m=m_e$  is the electron mass, and we indicated momenta of 
particles in the brackets.

\begin{figure}
\linethickness{0.5pt}
\begin{center}
\begin{picture}(62,77)(0,0)
\put(24.0,70.2){\vector(1,0){2}}
\put(9.9,70.2){\makebox(0,0)[r]{$e^-$}}
\put(10.9,70.2){\line(1,0){26.1}}
\put(43.5,70.2){\vector(1,0){2}}
\put(51.1,70.2){\makebox(0,0)[l]{$e^-$}}
\put(37.0,70.2){\line(1,0){13.0}}
\put(34.9,57.1){\makebox(0,0)[r]{$\gamma$}}
\multiput(36.7,70.2)(0.0,-0.3){8}{\rule[-0.25pt]{0.5pt}{0.5pt}}
\multiput(36.7,66.2)(0.0,-0.3){8}{\rule[-0.25pt]{0.5pt}{0.5pt}}
\multiput(36.7,62.2)(0.0,-0.3){8}{\rule[-0.25pt]{0.5pt}{0.5pt}}
\multiput(36.7,58.2)(0.0,-0.3){8}{\rule[-0.25pt]{0.5pt}{0.5pt}}
\multiput(36.7,54.2)(0.0,-0.3){8}{\rule[-0.25pt]{0.5pt}{0.5pt}}
\multiput(36.7,50.2)(0.0,-0.3){8}{\rule[-0.25pt]{0.5pt}{0.5pt}}
\multiput(36.7,46.2)(0.0,-0.3){8}{\rule[-0.25pt]{0.5pt}{0.5pt}}
\put(43.5,44.0){\vector(-1,0){2}}
\put(51.1,44.0){\makebox(0,0)[l]{$W^-$}}
\multiput(36.7,44.0)(0.3,0.0){12}{\rule[-0.25pt]{0.5pt}{0.5pt}}
\multiput(41.6,44.0)(0.3,0.0){12}{\rule[-0.25pt]{0.5pt}{0.5pt}}
\multiput(46.5,44.0)(0.3,0.0){12}{\rule[-0.25pt]{0.5pt}{0.5pt}}
\put(37.0,30.9){\vector(0,-1){2}}
\put(34.9,30.9){\makebox(0,0)[r]{$W^+$}}
\multiput(36.7,44.0)(0.0,-0.3){8}{\rule[-0.25pt]{0.5pt}{0.5pt}}
\multiput(36.7,40.0)(0.0,-0.3){8}{\rule[-0.25pt]{0.5pt}{0.5pt}}
\multiput(36.7,36.0)(0.0,-0.3){8}{\rule[-0.25pt]{0.5pt}{0.5pt}}
\multiput(36.7,32.0)(0.0,-0.3){8}{\rule[-0.25pt]{0.5pt}{0.5pt}}
\multiput(36.7,28.0)(0.0,-0.3){8}{\rule[-0.25pt]{0.5pt}{0.5pt}}
\multiput(36.7,24.0)(0.0,-0.3){8}{\rule[-0.25pt]{0.5pt}{0.5pt}}
\multiput(36.7,20.0)(0.0,-0.3){8}{\rule[-0.25pt]{0.5pt}{0.5pt}}
\put(9.9,17.8){\makebox(0,0)[r]{$\gamma$}}
\multiput(10.7,17.8)(0.3,0.0){8}{\rule[-0.25pt]{0.5pt}{0.5pt}}
\multiput(14.7,17.8)(0.3,0.0){8}{\rule[-0.25pt]{0.5pt}{0.5pt}}
\multiput(18.7,17.8)(0.3,0.0){8}{\rule[-0.25pt]{0.5pt}{0.5pt}}
\multiput(22.6,17.8)(0.3,0.0){8}{\rule[-0.25pt]{0.5pt}{0.5pt}}
\multiput(26.6,17.8)(0.3,0.0){8}{\rule[-0.25pt]{0.5pt}{0.5pt}}
\multiput(30.6,17.8)(0.3,0.0){8}{\rule[-0.25pt]{0.5pt}{0.5pt}}
\multiput(34.6,17.8)(0.3,0.0){8}{\rule[-0.25pt]{0.5pt}{0.5pt}}
\put(43.5,17.8){\vector(1,0){2}}
\put(51.1,17.8){\makebox(0,0)[l]{$W^+$}}
\multiput(36.7,17.8)(0.3,0.0){12}{\rule[-0.25pt]{0.5pt}{0.5pt}}
\multiput(41.6,17.8)(0.3,0.0){12}{\rule[-0.25pt]{0.5pt}{0.5pt}}
\multiput(46.5,17.8)(0.3,0.0){12}{\rule[-0.25pt]{0.5pt}{0.5pt}}
\end{picture} \hspace{1cm}
\begin{picture}(62,77)(0,0)
\put(24.0,70.2){\vector(1,0){2}}
\put(9.9,70.2){\makebox(0,0)[r]{$e^-$}}
\put(10.9,70.2){\line(1,0){26.1}}
\put(43.5,70.2){\vector(1,0){2}}
\put(51.1,70.2){\makebox(0,0)[l]{$e^-$}}
\put(37.0,70.2){\line(1,0){13.0}}
\put(34.9,57.1){\makebox(0,0)[r]{$\gamma$}}
\multiput(36.7,70.2)(0.0,-0.3){8}{\rule[-0.25pt]{0.5pt}{0.5pt}}
\multiput(36.7,66.2)(0.0,-0.3){8}{\rule[-0.25pt]{0.5pt}{0.5pt}}
\multiput(36.7,62.2)(0.0,-0.3){8}{\rule[-0.25pt]{0.5pt}{0.5pt}}
\multiput(36.7,58.2)(0.0,-0.3){8}{\rule[-0.25pt]{0.5pt}{0.5pt}}
\multiput(36.7,54.2)(0.0,-0.3){8}{\rule[-0.25pt]{0.5pt}{0.5pt}}
\multiput(36.7,50.2)(0.0,-0.3){8}{\rule[-0.25pt]{0.5pt}{0.5pt}}
\multiput(36.7,46.2)(0.0,-0.3){8}{\rule[-0.25pt]{0.5pt}{0.5pt}}
\put(43.5,44.0){\vector(1,0){2}}
\put(51.1,44.0){\makebox(0,0)[l]{$W^+$}}
\multiput(36.7,44.0)(0.3,0.0){12}{\rule[-0.25pt]{0.5pt}{0.5pt}}
\multiput(41.6,44.0)(0.3,0.0){12}{\rule[-0.25pt]{0.5pt}{0.5pt}}
\multiput(46.5,44.0)(0.3,0.0){12}{\rule[-0.25pt]{0.5pt}{0.5pt}}
\put(37.0,30.9){\vector(0,1){2}}
\put(34.9,30.9){\makebox(0,0)[r]{$W^+$}}
\multiput(36.7,44.0)(0.0,-0.3){8}{\rule[-0.25pt]{0.5pt}{0.5pt}}
\multiput(36.7,40.0)(0.0,-0.3){8}{\rule[-0.25pt]{0.5pt}{0.5pt}}
\multiput(36.7,36.0)(0.0,-0.3){8}{\rule[-0.25pt]{0.5pt}{0.5pt}}
\multiput(36.7,32.0)(0.0,-0.3){8}{\rule[-0.25pt]{0.5pt}{0.5pt}}
\multiput(36.7,28.0)(0.0,-0.3){8}{\rule[-0.25pt]{0.5pt}{0.5pt}}
\multiput(36.7,24.0)(0.0,-0.3){8}{\rule[-0.25pt]{0.5pt}{0.5pt}}
\multiput(36.7,20.0)(0.0,-0.3){8}{\rule[-0.25pt]{0.5pt}{0.5pt}}
\put(9.9,17.8){\makebox(0,0)[r]{$\gamma$}}
\multiput(10.7,17.8)(0.3,0.0){8}{\rule[-0.25pt]{0.5pt}{0.5pt}}
\multiput(14.7,17.8)(0.3,0.0){8}{\rule[-0.25pt]{0.5pt}{0.5pt}}
\multiput(18.7,17.8)(0.3,0.0){8}{\rule[-0.25pt]{0.5pt}{0.5pt}}
\multiput(22.6,17.8)(0.3,0.0){8}{\rule[-0.25pt]{0.5pt}{0.5pt}}
\multiput(26.6,17.8)(0.3,0.0){8}{\rule[-0.25pt]{0.5pt}{0.5pt}}
\multiput(30.6,17.8)(0.3,0.0){8}{\rule[-0.25pt]{0.5pt}{0.5pt}}
\multiput(34.6,17.8)(0.3,0.0){8}{\rule[-0.25pt]{0.5pt}{0.5pt}}
\put(43.5,17.8){\vector(-1,0){2}}
\put(51.1,17.8){\makebox(0,0)[l]{$W^-$}}
\multiput(36.7,17.8)(0.3,0.0){12}{\rule[-0.25pt]{0.5pt}{0.5pt}}
\multiput(41.6,17.8)(0.3,0.0){12}{\rule[-0.25pt]{0.5pt}{0.5pt}}
\multiput(46.5,17.8)(0.3,0.0){12}{\rule[-0.25pt]{0.5pt}{0.5pt}}
\end{picture} \hspace{1cm}
\begin{picture}(62,77)(0,0)
\put(24.0,57.1){\vector(1,0){2}}
\put(9.9,57.1){\makebox(0,0)[r]{$e^-$}}
\put(10.9,57.1){\line(1,0){26.1}}
\put(43.5,63.6){\vector(1,1){2}}
\put(51.1,70.2){\makebox(0,0)[l]{$e^-$}}
\put(37.0,57.1){\line(1,1){13.0}}
\put(34.9,44.0){\makebox(0,0)[r]{$\gamma$}}
\multiput(36.7,57.1)(0.0,-0.3){8}{\rule[-0.25pt]{0.5pt}{0.5pt}}
\multiput(36.7,53.1)(0.0,-0.3){8}{\rule[-0.25pt]{0.5pt}{0.5pt}}
\multiput(36.7,49.1)(0.0,-0.3){8}{\rule[-0.25pt]{0.5pt}{0.5pt}}
\multiput(36.7,45.1)(0.0,-0.3){8}{\rule[-0.25pt]{0.5pt}{0.5pt}}
\multiput(36.7,41.1)(0.0,-0.3){8}{\rule[-0.25pt]{0.5pt}{0.5pt}}
\multiput(36.7,37.1)(0.0,-0.3){8}{\rule[-0.25pt]{0.5pt}{0.5pt}}
\multiput(36.7,33.1)(0.0,-0.3){8}{\rule[-0.25pt]{0.5pt}{0.5pt}}
\put(9.9,30.9){\makebox(0,0)[r]{$\gamma$}}
\multiput(10.7,30.9)(0.3,0.0){8}{\rule[-0.25pt]{0.5pt}{0.5pt}}
\multiput(14.7,30.9)(0.3,0.0){8}{\rule[-0.25pt]{0.5pt}{0.5pt}}
\multiput(18.7,30.9)(0.3,0.0){8}{\rule[-0.25pt]{0.5pt}{0.5pt}}
\multiput(22.6,30.9)(0.3,0.0){8}{\rule[-0.25pt]{0.5pt}{0.5pt}}
\multiput(26.6,30.9)(0.3,0.0){8}{\rule[-0.25pt]{0.5pt}{0.5pt}}
\multiput(30.6,30.9)(0.3,0.0){8}{\rule[-0.25pt]{0.5pt}{0.5pt}}
\multiput(34.6,30.9)(0.3,0.0){8}{\rule[-0.25pt]{0.5pt}{0.5pt}}
\put(43.5,37.5){\vector(1,1){2}}
\put(51.1,44.0){\makebox(0,0)[l]{$W^+$}}
\multiput(36.7,30.9)(0.3,0.3){12}{\rule[-0.25pt]{0.5pt}{0.5pt}}
\multiput(41.6,35.8)(0.3,0.3){12}{\rule[-0.25pt]{0.5pt}{0.5pt}}
\multiput(46.5,40.7)(0.3,0.3){12}{\rule[-0.25pt]{0.5pt}{0.5pt}}
\put(43.5,24.4){\vector(-1,1){2}}
\put(51.1,17.8){\makebox(0,0)[l]{$W^-$}}
\multiput(36.7,30.9)(0.3,-0.3){12}{\rule[-0.25pt]{0.5pt}{0.5pt}}
\multiput(41.6,26.0)(0.3,-0.3){12}{\rule[-0.25pt]{0.5pt}{0.5pt}}
\multiput(46.5,21.1)(0.3,-0.3){12}{\rule[-0.25pt]{0.5pt}{0.5pt}}
\end{picture}
\end{center}
\caption{Feynman diagrams of $e^-+\gamma\to e^-+ W^++W^-$ with
$t$-channel type of singularity, produced by the exchange of
virtual photon.}\label{fig:1a}
\end{figure}
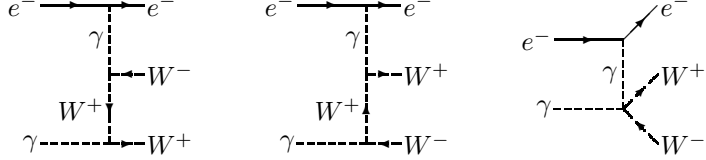

Another example is given by the process
$$e^-(p_1)+\gamma(p_2)\rightarrow e^-(p_3)+W^+(p_4)+W^-(p_5)\,,$$
when {\it in\/}-electron radiates the virtual photon (here
$m=0$). The Feynman diagrams for this case are shown in
Fig.~\ref{fig:1a}.

Separate squared diagrams can include terms asymptotically
proportional to $1/t^2$ and also to $1/t$.  The $1/t^2$ terms
coming from different diagrams can cancel each other up to the
terms of $1/t$ type (in the $m_e\to 0$ limit) due to the gauge invariance.  
However, for processes $2\to n$, with $n>2$, if the
singular exchange propagator has its momentum composed from more
than one out-particle momenta then there is no cancellation of
$1/t^2$ terms. For some processes of $2\to 2$ type, 
like elastic scattering of two charged particles, there are no
cancellation of $1/t^2$ terms also. 
In these cases we have much stronger singularity.

Let us consider a $t$-channel singularity over $t_I=(p^{\rm
in}-\sum p_i^{\rm out})^2$, where $p^{\rm in}$ is momentum of one of
in-particles and sum is made over the some subset of out-particles. 
Suppose that we constructed a tree of 
{\it elementary decays} in such a way that
we can choose some {\it elementary decay\/}, with 
clusters momenta $q_1$ and $q_{2}$, 
and the reference vector ${\cal P}$, as a combination of clusters momenta
from previous {\it elementary decays},
according to one of two variants:
\begin{equation}	
\left(p^{\rm in}-\sum p_i^{\rm out}\right)^2=(q_1+{\cal P})^2\,,
\quad\mbox{or}\quad
\left(p^{\rm in}-\sum p_i^{\rm out}\right)^2=(q_{2}+{\cal P})^2\,.
\end{equation}
In these cases we have
\begin{equation}	
t_I=(q)^2+({\cal P})^2+2(q_0{\cal P}_0-|\vec{q}|
|{\bar {\cal P}}|\cos\Theta)\,,\quad q=q_{1,2}\,.
\end{equation}
So, the invariant $t_I$ will be linear function of the
integration variable $\cos\Theta$.

In the case of cancellation of $1/t^2$ terms
the model singular function is
\begin{equation}	
g_i(\cos\Theta)=\frac1{\cos\Theta-C_0}\,.
\end{equation}
Its integral function is
\begin{equation}	
G_i(\cos\Theta)=\ln|\cos\Theta-C_0|\,,
\end{equation}
where
$$C_0=\frac{m^2-({\cal P})^2-(q)^2+2q_0
{\cal P}_0}{2|{\bar {\cal P}}||\vec{q}|}\,.$$
Here $m^2$ is the squared mass of the corresponding singular
propagator (Eq.~(\ref{eq:tprop})).

In the cases without cancellation of $1/t^2$
terms  the model singular function is
\begin{equation}	
g_i(\cos\Theta)=\frac1{(\cos\Theta-C_0)^2}\,,
\end{equation}
with integral function
\begin{equation}	
G_i(\cos\Theta)=\frac1{C_0-\cos\Theta}\,.
\end{equation}

\subsubsection{$s$-channel decay pole}
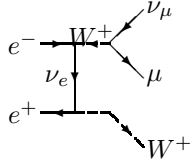
\begin{figure}
\linethickness{0.5pt}
\begin{center}
\begin{picture}(62,77)(0,0)
\put(17.5,57.1){\vector(1,0){2}}
\put(9.9,57.1){\makebox(0,0)[r]{$e^-$}}
\put(10.9,57.1){\line(1,0){13.0}}
\put(30.5,57.1){\vector(-1,0){2}}
\put(30.2,60.2){\makebox(0,0){$W^+$}}
\multiput(23.7,57.1)(0.3,0.0){12}{\rule[-0.25pt]{0.5pt}{0.5pt}}
\multiput(28.6,57.1)(0.3,0.0){12}{\rule[-0.25pt]{0.5pt}{0.5pt}}
\multiput(33.5,57.1)(0.3,0.0){12}{\rule[-0.25pt]{0.5pt}{0.5pt}}
\put(43.5,63.6){\vector(-1,-1){2}}
\put(51.1,70.2){\makebox(0,0)[l]{$\bar\nu_\mu$}}
\put(37.0,57.1){\line(1,1){13.0}}
\put(43.5,50.5){\vector(1,-1){2}}
\put(51.1,44.0){\makebox(0,0)[l]{$\mu$}}
\put(37.0,57.1){\line(1,-1){13.0}}
\put(24.0,44.0){\vector(0,-1){2}}
\put(21.9,44.0){\makebox(0,0)[r]{$\nu_e$}}
\put(24.0,57.1){\line(0,-1){26.2}}
\put(17.5,30.9){\vector(-1,0){2}}
\put(9.9,30.9){\makebox(0,0)[r]{$e^+$}}
\put(10.9,30.9){\line(1,0){13.0}}
\multiput(23.7,30.9)(0.3,0.0){12}{\rule[-0.25pt]{0.5pt}{0.5pt}}
\multiput(28.6,30.9)(0.3,0.0){12}{\rule[-0.25pt]{0.5pt}{0.5pt}}
\multiput(33.5,30.9)(0.3,0.0){12}{\rule[-0.25pt]{0.5pt}{0.5pt}}
\put(43.5,24.4){\vector(1,-1){2}}
\put(51.1,17.8){\makebox(0,0)[l]{$W^+$}}
\multiput(36.7,30.9)(0.3,-0.3){12}{\rule[-0.25pt]{0.5pt}{0.5pt}}
\multiput(41.6,26.0)(0.3,-0.3){12}{\rule[-0.25pt]{0.5pt}{0.5pt}}
\multiput(46.5,21.1)(0.3,-0.3){12}{\rule[-0.25pt]{0.5pt}{0.5pt}}
\end{picture}
\end{center}
\caption{Feynman diagram with decayed $W$ propagator
contributed in the process $e^-+e^+\to\bar\nu_\mu+\mu+W^+$.}
\label{fig:diag2}
\end{figure}

\noindent
When a process includes the $s$-channel decay of some unstable
particle it is necessary to introduce a finite width in the
corresponding propagator to get a physically meaningful (and
finite) cross section. Such diagrams will give in squared matrix
element the factor
\begin{equation}	
\frac1{(s_k-M^2)^2+\Gamma^2\cdot M^2}\,,
\end{equation}
where $M$ is the mass of unstable particle, $\Gamma$ is its
width and $s_k$ is the squared invariant mass of decay products.
The integrand has a sharp peak if
\begin{equation}	
\Gamma\cdot M\ll s\,.
\end{equation}
An example is the process
$$e^-+e^+\rightarrow\bar\nu_\mu+\mu+W^+$$
which includes the $W^-$ decay into the lepton pair
$\bar\nu_\mu\mu$.  The Feynman diagram with such singularity is
shown in Fig.~\ref{fig:diag2}.

If $s_k$ is an integration variable (so it is squared decayed
momentum of one {\it elementary decay\/} of the kinematical
scheme) the corresponding model singular function can be taken as
\begin{equation}	
g_i(s_k)=\frac1{(s_k-M^2)^2+\varepsilon^2}
\end{equation}
and its integral function equals
\begin{equation}	
G_i(s_k)=\frac1{\varepsilon}\arctan
\left(\frac{s_k-M^2}{\varepsilon}\right)\,,
\end{equation}
where $\varepsilon\equiv \Gamma\cdot M$.

Let us now consider the case when in some kinematical scheme
$s_k$ does not correspond to decayed momentum of any {\it
elementary decay.} So we can not regularize this singularity by
the previous way.  To carry out the regularization in this case
we need to choose {\it elementary decay}, with that 
4-momenta $q_1$ and $q_2$ of its two clusters, and
the reference vector ${\cal P}$, as a combination of clusters momenta from
previous {\it elementary decays},  according to
one of the following variants:
\begin{equation}	
s_k=\left(\sum p_i^{\rm out}\right)^2=(q_1+{\cal P})^2\,,\quad\mbox{or}\quad
s_k=\left(\sum p_i^{\rm out}\right)^2=(q_2+{\cal P})^2\,.
\end{equation}
In these cases we have
\begin{equation}	
s_k=(q)^2+({\cal P})^2+2(q_0{\cal P}_0-|\vec{q}||
{\bar {\cal P}}|\cos\Theta)\,,\quad q=q_{1,2}\,.
\label{eq:s-chan-theta}
\end{equation}
Thus the invariant $s_k$ will be linear function of the
integration variable $\cos\Theta$.  The model singular function
for such cases is
\begin{equation}	
g_i(\cos\Theta)=\frac1{(\cos\Theta-C_0)^2+\varepsilon_0^2}
\end{equation}
and its integral function equals
\begin{equation}	
G_i(\cos\Theta)=\frac1{\varepsilon_0}\arctan
\left(\frac{\cos\Theta-C_0}{\varepsilon_0}\right)\,,
\end{equation}
where
$$C_0=\frac{M^2-(q)^2-({\cal P})^2-2 q_0{\cal P}_0}
{-2|\vec{q}||{\bar {\cal P}}|}\,,$$
and
$$\varepsilon_0=\frac{\varepsilon}{-2|\vec{q}||{\bar {\cal P}}|}\,.$$

\subsubsection{Infrared type of the singularity}
\begin{figure}
\linethickness{0.5pt}
\begin{center}
\begin{picture}(62,77)(0,0)
\put(17.5,57.1){\vector(1,0){2}}
\put(9.9,57.1){\makebox(0,0)[r]{$e^-$}}
\put(10.9,57.1){\line(1,0){13.0}}
\put(30.2,60.2){\makebox(0,0){$\gamma$}}
\multiput(23.7,57.1)(0.3,0.0){12}{\rule[-0.25pt]{0.5pt}{0.5pt}}
\multiput(28.6,57.1)(0.3,0.0){12}{\rule[-0.25pt]{0.5pt}{0.5pt}}
\multiput(33.5,57.1)(0.3,0.0){12}{\rule[-0.25pt]{0.5pt}{0.5pt}}
\put(43.5,63.6){\vector(1,1){2}}
\put(51.1,70.2){\makebox(0,0)[l]{$e^-$}}
\put(37.0,57.1){\line(1,1){13.0}}
\put(43.5,50.5){\vector(-1,1){2}}
\put(51.1,44.0){\makebox(0,0)[l]{$e^+$}}
\put(37.0,57.1){\line(1,-1){13.0}}
\put(24.0,44.0){\vector(0,-1){2}}
\put(21.9,44.0){\makebox(0,0)[r]{$e^-$}}
\put(24.0,57.1){\line(0,-1){26.2}}
\put(17.5,30.9){\vector(-1,0){2}}
\put(9.9,30.9){\makebox(0,0)[r]{$e^+$}}
\put(10.9,30.9){\line(1,0){13.0}}
\multiput(23.7,30.9)(0.3,0.0){12}{\rule[-0.25pt]{0.5pt}{0.5pt}}
\multiput(28.6,30.9)(0.3,0.0){12}{\rule[-0.25pt]{0.5pt}{0.5pt}}
\multiput(33.5,30.9)(0.3,0.0){12}{\rule[-0.25pt]{0.5pt}{0.5pt}}
\put(51.1,17.8){\makebox(0,0)[l]{$Z$}}
\multiput(36.7,30.9)(0.3,-0.3){12}{\rule[-0.25pt]{0.5pt}{0.5pt}}
\multiput(41.6,26.0)(0.3,-0.3){12}{\rule[-0.25pt]{0.5pt}{0.5pt}}
\multiput(46.5,21.1)(0.3,-0.3){12}{\rule[-0.25pt]{0.5pt}{0.5pt}}
\end{picture} \hspace{1cm}
\begin{picture}(62,77)(0,0)
\put(17.5,57.1){\vector(1,0){2}}
\put(9.9,57.1){\makebox(0,0)[r]{$e^-$}}
\put(10.9,57.1){\line(1,0){13.0}}
\multiput(23.7,57.1)(0.3,0.0){12}{\rule[-0.25pt]{0.5pt}{0.5pt}}
\multiput(28.6,57.1)(0.3,0.0){12}{\rule[-0.25pt]{0.5pt}{0.5pt}}
\multiput(33.5,57.1)(0.3,0.0){12}{\rule[-0.25pt]{0.5pt}{0.5pt}}
\put(51.1,70.2){\makebox(0,0)[l]{$Z$}}
\multiput(36.7,57.1)(0.3,0.3){12}{\rule[-0.25pt]{0.5pt}{0.5pt}}
\multiput(41.6,62.0)(0.3,0.3){12}{\rule[-0.25pt]{0.5pt}{0.5pt}}
\multiput(46.5,66.9)(0.3,0.3){12}{\rule[-0.25pt]{0.5pt}{0.5pt}}
\put(24.0,44.0){\vector(0,-1){2}}
\put(21.9,44.0){\makebox(0,0)[r]{$e^-$}}
\put(24.0,57.1){\line(0,-1){26.2}}
\put(17.5,30.9){\vector(-1,0){2}}
\put(9.9,30.9){\makebox(0,0)[r]{$e^+$}}
\put(10.9,30.9){\line(1,0){13.0}}
\put(30.2,34.0){\makebox(0,0){$\gamma$}}
\multiput(23.7,30.9)(0.3,0.0){12}{\rule[-0.25pt]{0.5pt}{0.5pt}}
\multiput(28.6,30.9)(0.3,0.0){12}{\rule[-0.25pt]{0.5pt}{0.5pt}}
\multiput(33.5,30.9)(0.3,0.0){12}{\rule[-0.25pt]{0.5pt}{0.5pt}}
\put(43.5,37.5){\vector(1,1){2}}
\put(51.1,44.0){\makebox(0,0)[l]{$e^-$}}
\put(37.0,30.9){\line(1,1){13.0}}
\put(43.5,24.4){\vector(-1,1){2}}
\put(51.1,17.8){\makebox(0,0)[l]{$e^+$}}
\put(37.0,30.9){\line(1,-1){13.0}}
\end{picture}
\end{center}
\caption{Feynman diagrams for $e^-+e^+\to e^-+e^++Z$ with
infrared type of kinematical singularity.}\label{fig:diag3}
\end{figure}
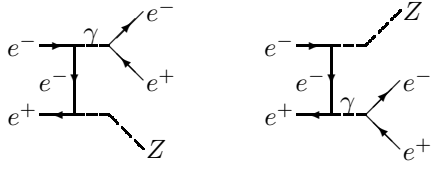

\noindent
Let two light out-particles with masses $m_1$ and $m_2$ (for
instance photon, electron, {\sf u} or {\sf d} quarks) be {\it
``products of the decay"\/} of some virtual particle with mass $m$,
and $m<m_1+m_2$. Sharp peak arises from the propagator of the
``decaying" virtual particle with invariant mass $s_i$:
\begin{equation}	
\frac1{(s_i-m^2)^2}\,, \quad s_i\geq(m_1+m_2)^2\,,
\end{equation}
if the masses $m_1$, $m_2$, and $m$ are small:
\begin{equation}	
m_1,\, m_2,\, m\ll\sqrt s\,.
\end{equation}
Note that in the Standard Model such singular structure appears
typically if one of these three particles is photon or gluon.
An example is a ``decay" of virtual photon into two light
charged particles. Such peak arises in some Feynman diagrams
(see Fig.~\ref{fig:diag3}) of the process
$$e^-+e^+\rightarrow e^-+e^++{\sf Z}\,.$$
In this case $m=0$ and $m_1=m_2=m_e$ --- mass of electron.

The singularity of infrared type is present also in the case of
out-photon emission from the charged out-particle. In this case
$m_1=0$ and $m_2=m$ is a mass of charged particle.  The total
cross section is infinite in this case --- there is a
nonintegrable pole at zero photon energy (so-called infrared
catastrophe).  However one can introduce the cut
\begin{equation}	
s_i>\delta^2+m^2\,, \quad s_i=(p_{m_1}+p_{m_2})^2\,,
\end{equation}
where $p_{m_1}$ and $p_{m_2}$ are momenta of the out-photon and
charged out-particle.  In this case sharp peak appears if
\begin{equation}	
\delta\ll\sqrt s\,.
\end{equation}
All these situations are different from the previous example of
$s$-channel singularity.  First, the pole is placed beyond the
physical region. Second, due to the gauge invariance, the
singularity will be of the first order, terms of $1/(s_i-m^2)^2$
type will always cancel each other up to the terms of
$1/(s_i-m^2)$ type.  So we have the similar functional problem
as in the case of $t$-channel pole and we can use the formulas
of the $t$-channel regularization to handle such poles.

\section{Computational Examples}
\noindent
The method described in the previous sections was realized in
CompHEP \cite{2} package. Let us inspect the efficiency of this
method in some examples. The results of cross section
calculations by CompHEP with Monte Carlo integration are listed
below. The program VEGAS \cite{8} was used for multidimensional
adaptive MC integration.

\subsection{{\it t}-channel singularity}
\noindent
Let us begin from the process
$$e^-(p_1)+\gamma(p_2)\rightarrow e^-(p_3)+Z(p_4)+H(p_5)$$
with singularity appearing from the exchange of electron in
$t$-channel (Fig.~\ref{fig:diag1}).
First let us use the default kinematics in CompHEP:
$$\begin{array}{ll}
\mbox{1st {\it elementary decay\/}:} 
&\quad\mbox{$p_1+p_2\rightarrow p_3,\, p_4+p_5$ with ${\cal P}=p_1$}\,;\\
\mbox{2nd {\it elementary decay\/}:}
&\quad\mbox{$p_4+p_5\rightarrow p_4,\, p_5$ with ${\cal P}=-p_1+p_3$}\,.
\end{array}$$
For the number of points per iteration equals 10000, we can see
the following CompHEP screen output for the total cross section
calculation at $\sqrt s=500$~GeV:
{\footnotesize
\begin{verbatim}
==============================================================================
  IT      CURRENT RESULTS                ACCUMULATED RESULTS        CHI**2
       Cross section [pb]  Error %    Cross section [pb]  Error %
------------------------------------------------------------------------------
   1   3.685360E-04        24.68      3.685360E-04        24.68     .00
   2   6.255344E-04        27.15      4.258167E-04        18.83    1.78
   3   4.638301E-04        10.09      4.541736E-04         8.90     .97
   4   5.303054E-04         6.19      5.000283E-04         5.10    1.36
   5   6.214947E-04        16.34      5.072243E-04         4.87    1.36
   6   6.355495E-04         8.07      5.314256E-04         4.19    2.10
   7   7.458411E-04         7.14      5.633162E-04         3.65    4.05
   8   9.273354E-04        19.66      5.678777E-04         3.59    4.04
   9   8.410235E-04         6.03      6.059285E-04         3.13    6.65
  10   7.566958E-04         1.20      7.285007E-04         1.12   11.64
______________________________________________________________________________
\end{verbatim}
}
Here IT is iteration number, CURRENT RESULTS are results of the
integration for current iteration. ACCUMULATED RESULTS are
cumulative value of the integration of current and previous
iterations.
These numbers , as well as the error and parameter $\chi^2$, are evaluated
according standard formulas (see details in Ref.~\cite{8}).

The result above shows that there is no stability from iteration
to iteration (see 2nd column --- values of current results; and
3rd column --- current errors). Also there is no reliability
with due to large value of $\chi^2$ parameter.

Let us now introduce the kinematical regularization of the
$t$-channel type for the invariant $(p_3-p_2)^2$ with mass
$m=m_e$.  Used kinematical scheme allows to introduce this
regularization because ${\cal P}$ in the 1st {\it elementary
decay\/} equals $p_1$ and $p_1=-p_2$ in the 1st decay.  The
CompHEP screen output of the integration with entered
regularization will be:
{\footnotesize
\begin{verbatim}
==============================================================================
  IT      CURRENT RESULTS                ACCUMULATED RESULTS        CHI**2
       Cross section [pb]  Error %    Cross section [pb]  Error %
------------------------------------------------------------------------------
   1   7.621777E-04          .33      7.621777E-04          .33     .00
   2   7.598378E-04          .21      7.605149E-04          .18     .62
   3   7.590317E-04          .18      7.597725E-04          .13     .61
______________________________________________________________________________
\end{verbatim}
}

\subsection{{\it s}-channel singularity}
\noindent
Let us consider an example with $s$-channel singularity (see
Fig.~\ref{fig:diag2}):
$$e^-(p_1)+e^+(p_2)\rightarrow\bar\nu_\mu(p_3)
+\mu(p_4)+W^+(p_5)\,.$$

If we use the following kinematical scheme
$$\begin{array}{ll}
\mbox{1st {\it elementary decay\/}:}
&\quad\mbox{$p_1+p_2\rightarrow p_3,\, p_4+p_5$ with ${\cal P}=p_2$}\,;\\
\mbox{2nd {\it elementary decay\/}:}
&\quad\mbox{$p_4+p_5\rightarrow p_4,\, p_5$ with ${\cal P}=p_3$}\,,
\end{array}$$
then the CompHEP screen output with 25000 points per iteration
and $\sqrt s=500$~GeV is:
{\footnotesize
\begin{verbatim}
==============================================================================
  IT      CURRENT RESULTS                ACCUMULATED RESULTS        CHI**2
       Cross section [pb]  Error %    Cross section [pb]  Error %
------------------------------------------------------------------------------
   1   6.407551E-01        27.60      6.407551E-01        27.60     .00
   2   8.139560E-01        10.75      7.798860E-01        10.06     .77
   3   7.667456E-01         3.48      7.681075E-01         3.29     .40
   4   8.065722E-01         2.70      7.901984E-01         2.08     .71
   5   8.483987E-01         4.06      8.010476E-01         1.86    1.11
   6   8.442694E-01         2.95      8.123698E-01         1.57    1.33
   7   8.276117E-01         2.50      8.165697E-01         1.33    1.18
   8   8.402845E-01         2.54      8.214562E-01         1.18    1.15
   9   8.512594E-01         3.52      8.242752E-01         1.12    1.12
  10   8.232579E-01         3.04      8.241540E-01         1.05     .99
______________________________________________________________________________
\end{verbatim}
}
We see that the error of current iteration (see 3rd column) is
standing at the level about 3\% after 10 iterations that it is
not satisfactory in many applications.

Let us choose other kinematical scheme
$$\begin{array}{ll}
\mbox{1st {\it elementary decay\/}:}
&\quad\mbox{$p_1+p_2\rightarrow p_3+p_4,\, p_5$ with ${\cal P}=p_2$}\\
\mbox{2nd {\it elementary decay\/}:}
&\quad\mbox{$p_3+p_4\rightarrow p_3,\, p_4$ with ${\cal P}=p_1$}\,.
\end{array}$$
One can see much better results even without regularization:
{\footnotesize
\begin{verbatim}
==============================================================================
  IT      CURRENT RESULTS                ACCUMULATED RESULTS        CHI**2
       Cross section [pb]  Error %    Cross section [pb]  Error %
------------------------------------------------------------------------------
   1   9.342104E-01        29.08      9.342104E-01        29.08     .00
   2   8.762828E-01         8.78      8.805859E-01         8.41     .04
   3   8.677625E-01         1.18      8.680033E-01         1.17     .04
   4   8.463662E-01          .65      8.513031E-01          .57    1.19
   5   8.471336E-01          .57      8.492184E-01          .40     .99
______________________________________________________________________________
\end{verbatim}
}
Then introducing regularization over invariant mass
$(p_3+p_4)^2$ with the mass parameter $M=M_W$ ($W$-boson mass)
and the width parameter $\Gamma=\Gamma_W$ ($W$-boson width) we
get more effective integration than in the previous case:
{\footnotesize
\begin{verbatim}
==============================================================================
  IT      CURRENT RESULTS                ACCUMULATED RESULTS        CHI**2
       Cross section [pb]  Error %    Cross section [pb]  Error %
------------------------------------------------------------------------------
   1   8.535685E-01         4.72      8.535685E-01         4.72     .00
   2   8.539195E-01          .94      8.539060E-01          .93     .00
   3   8.544500E-01          .60      8.542897E-01          .50     .00
   4   8.449297E-01          .58      8.502279E-01          .38     .69
   5   8.481124E-01          .57      8.495686E-01          .32     .55
______________________________________________________________________________
\end{verbatim}
}
The result of regularization is strong decrease of the error of
first iterations.

However, it can happen that for some reason one cannot choose
kinematical scheme where this $s$-channel singularity is related
to some integration variable as an invariant mass of {\it
elementary decay.} In the process $e^-+e^+\to
e^-+\bar\nu_e+W^+$, for example, when discussed $s$-channel
singularity contributes, other singularities are present and
they can force us to introduce another kinematics to regularize
these peaks. In such cases one can use regularization of the
$s$-channel singularity using polar angle $\Theta$
(Eq.~(\ref{eq:s-chan-theta})).  In our demonstration calculation
let us return to the first kinematical scheme given at the
beginning of this section and introduce the regularization for
the corresponding $(p_3+p_4)^2$ invariant.  It is possible to do
so because $p_3$ has been choosen in this kinematical scheme as
reference vector ${\cal P}$ in the 2nd {\it elementary decay.}
The CompHEP screen output for this case will be:
{\footnotesize
\begin{verbatim}
==============================================================================
  IT      CURRENT RESULTS                ACCUMULATED RESULTS        CHI**2
       Cross section [pb]  Error %    Cross section [pb]  Error %
------------------------------------------------------------------------------
   1   8.854719E-01         4.85      8.854719E-01         4.85     .00
   2   8.453821E-01         1.00      8.468828E-01          .98     .84
   3   8.477949E-01          .77      8.474495E-01          .60     .42
   4   8.497170E-01          .82      8.482457E-01          .49     .30
   5   8.503980E-01          .75      8.488813E-01          .41     .25
______________________________________________________________________________
\end{verbatim}
}
One has to compare this screen output with the first one at the
beginning of this section.  We see that the integration is
stable with fast convergency, smaller error and reliable value
of $\chi^2$ parameter, similar to the previous case when the
regularization over $(p_3+p_4)^2$ as invariant mass of {\it
elementary decay\/} was performed.

\section*{Acknowledgments}
\noindent
This work was partially supported by European Association INTAS
(grant 93-1180-ext), Russian Foundation for Basic Reseach (RFBR grant
96-02-19773a) and JSPS (Japan Society for Promotion Science).

We are grateful to E.~E.~Boos, M.~N.~Dubinin, and
S.~A.~Shichanin for fruitful discussions and testing the
program, and to Y.~Shimizu for useful discussions.


\end{document}